\documentclass{Interspeech}
\ninept

\interspeechcameraready

\title{Towards Temporally Explainable Dysarthric Speech Clarity Assessment}

\author[affiliation={2}]{Seohyun}{Park$^*$}
\author[affiliation={1}]{Chitralekha}{Gupta$^*$}
\author[affiliation={3}]{Michelle}{Kah Yian Kwan}
\author[affiliation={3}]{Xinhui}{Fung}
\author[affiliation={3}]{Alexander Wenjun}{Yip}
\author[affiliation={1}]{Suranga}{Nanayakkara}

\affiliation{Augmented Human Lab, School of Computing}{National University of Singapore}{Singapore}
\affiliation{}{Department of Data Science, Korea University}{Republic of Korea}
\affiliation{}{Department of Healthcare Redesign, Alexandra Hospital}{Singapore}
\email{emily21@korea.ac.kr, chitralekha@nus.edu.sg, Kah\_Yian\_Kwan@nuhs.edu.sg, xinhui\_fung@nuhs.edu.sg, Alexander\_yip@nuhs.edu.sg, suranga@comp.nus.edu.sg}


\usepackage{comment}
\usepackage{tabularx}
\usepackage[hang,flushmargin]{footmisc}
\begin{document}

\maketitle

\begin{abstract}
    

Dysarthria, a motor speech disorder, affects intelligibility and requires targeted interventions for effective communication. 
In this work, we investigate automated mispronunciation feedback by collecting a dysarthric speech dataset from six speakers reading two passages, annotated by a speech therapist with temporal markers and mispronunciation descriptions. We design a three-stage framework for explainable mispronunciation evaluation: (1) overall clarity scoring, (2) mispronunciation localization, and (3) mispronunciation type classification. We systematically analyze pretrained Automatic Speech Recognition (ASR) models in each stage, assessing their effectiveness in dysarthric speech evaluation\footnote{Code available at: \url{https://github.com/augmented-human-lab/interspeech25_speechtherapy}\\\noindent\textbf{$^*$Equal contributions by the first two authors}}. Our findings offer clinically relevant insights for automating actionable feedback for pronunciation assessment, which could enable independent practice for patients and help therapists deliver more effective interventions.

\end{abstract}

\section{Introduction}
Dysarthria, a speech disorder caused by neuromuscular impairments, affects articulation, phonation, and prosody, leading to communication barriers. Reliable assessment tools are needed to support personalized speech therapy. Current assessment methods rely heavily on expert judgment, which is time-consuming, limited in scalability, and with infrequent access. Automated evaluation systems based on ASR models have the potential to offer objective and scalable assessments.

ASR models have been explored for dysarthric speech recognition \cite{kim2018dysarthric,wang2024enhancing,leung2024training,lee2025dypcl}. Kim et al.~\cite{kim2018dysarthric} developed a CNN-LSTM system to improve recognition accuracy, while Wang et al.~\cite{wang2024enhancing} designed a prototype-based adaptation method for better performance without fine-tuning. 
This distinction is critical: unlike ASR systems that aim to correct errors for better transcription, assessment system must instead detect and highlight unintelligible regions to provide actionable feedback.

Several studies have focused on the overall speaker intelligibility assessment for individuals with dysarthria. 
For e.g., Rathod et al.~\cite{rathod2023whisper} proposed a transfer learning approach using features of the Whisper model's encoder, achieving higher accuracy in severity levels of dysarthria compared to traditional features such as MFCC. 
However, these approaches lack granularity in the feedback and only provide an overall assessment of speech intelligibility. Techniques such as \textit{goodness of pronunciation} have been used for phoneme-level evaluation in dysarthric speech \cite{yeo2023speech, choi2025leveraging}, but these methods are unable to assess their accuracy in pinpointing temporal regions of reduced intelligibility or in classifying dysarthria-specific mispronunciation types. One key reason for the limited research in this area is the scarcity of dysarthric speech datasets that include detailed annotations.

Public datasets for research in speech disorders are limited  \cite{rudzicz2012torgo,kim2008dysarthric,turrisi2021easycall}. The TORGO \cite{rudzicz2012torgo} and UA-Speech \cite{kim2008dysarthric} datasets provide dysarthric speech samples with varying severity or intelligibility levels, including isolated words, and sentences, which serve as valuable benchmarks for ASR evaluation under dysarthric conditions. However, these datasets lack fine-grained temporal annotations that mark specific regions of reduced intelligibility and detailed descriptions of mispronunciation within utterances. As a result, these limitations hinder their effectiveness in training and validating automated systems, especially for detailed error localization and articulation analysis.

Recent advancements in mispronunciation detection frameworks, such as those using the L2-Arctic corpus \cite{zhao18b_interspeech}, focus on accent correction and pronunciation errors in non-native speakers \cite{peng2021study,yan2021end}. However, these frameworks are not suitable for dysarthric speech, which prioritizes improving intelligibility over accent correction. Dysarthric speech is characterized by distinct articulatory and phonological impairments, which necessitate targeted interventions. These challenges underscore the need for datasets developed in collaboration with speech therapists, tailored to the unique characteristics of dysarthric speech. These datasets would allow automated systems to deliver precise, clinically relevant feedback, helping patients practice independently and enhance their speech intelligibility. 

In this work, we develop a three-stage framework that assigns clarity scores at the utterance level, localizes mispronounced regions, and classifies specific mispronunciation types. We collected a dysarthric speech dataset from six patients reading two English passages, annotated with temporal markers and mispronunciation descriptions by an expert speech therapist. We conduct an in-depth analysis of the performance of sota ASR models across the three stages, analyzing their strengths and limitations in dysarthric speech analysis. Our contributions are: (1) analysis of characteristics of dysarthric speech relevant to speech intelligibility as identified by an expert speech therapist, and (2) insights into the effectiveness of pre-trained ASR models for mispronunciation localization and classification, enabling fine-grained dysarthric speech assessment. 

\vspace{-0.3cm}
\section{Dataset Design}
Our dataset comprises 12 speech recordings with a mean speech duration of 1.06 min per recording, from six patients aged 41 to 71 years, diagnosed with various types of dysarthria, such as unilateral upper motor neuron, hyperkinetic, hypokinetic, and ataxic. The severity of their speech intelligibility ranged from mild to moderate, as assessed by the speech therapists. Patients were recruited by therapists, and informed consent was obtained. Each participant read two English passages: \textit{The Rainbow Passage}\footnote{\url{https://www.rit.edu/ntid/slpros/media/rainbow}} and \textit{The Grandfather Story}\footnote{\url{https://eatspeakthink.com/wp-content/uploads/2020/04/Grandfather-passage.pdf}}. The recordings were anonymized to protect patient privacy.

Errors that affect intelligibility were identified and annotated by a speech therapist specialized in neurorehabilitation with 9 years of experience. Annotated errors align with those commonly addressed in dysarthria therapy and reflect clinical practice. 
While some variability in error marking is expected due to the subjective nature of intelligibility assessment \cite{hirsch2022reliability}, experienced therapists typically annotate these errors consistently. Additionally, local accent variations were carefully noted and not marked as errors, as they are part of regional speech patterns and do not affect intelligibility.


 Our dataset's speaker diversity is comparable to existing dysarthric speech datasets; for e.g., the TORGO dataset includes eight speakers across four severity levels, while ours comprises six speakers across two levels. However, the size of our dataset is limited because obtaining carefully marked clinically relevant annotations from speech therapists is highly time-intensive and requires specialized expertise. 
 Although modest in size, our dataset is unique in providing temporally aligned, therapist-labeled mispronunciation annotations—resources not available in existing datasets yet essential for developing automated assessment tools.
 
\begin{figure}
    \vspace{-0.3cm}\centering\includegraphics[width=\columnwidth]{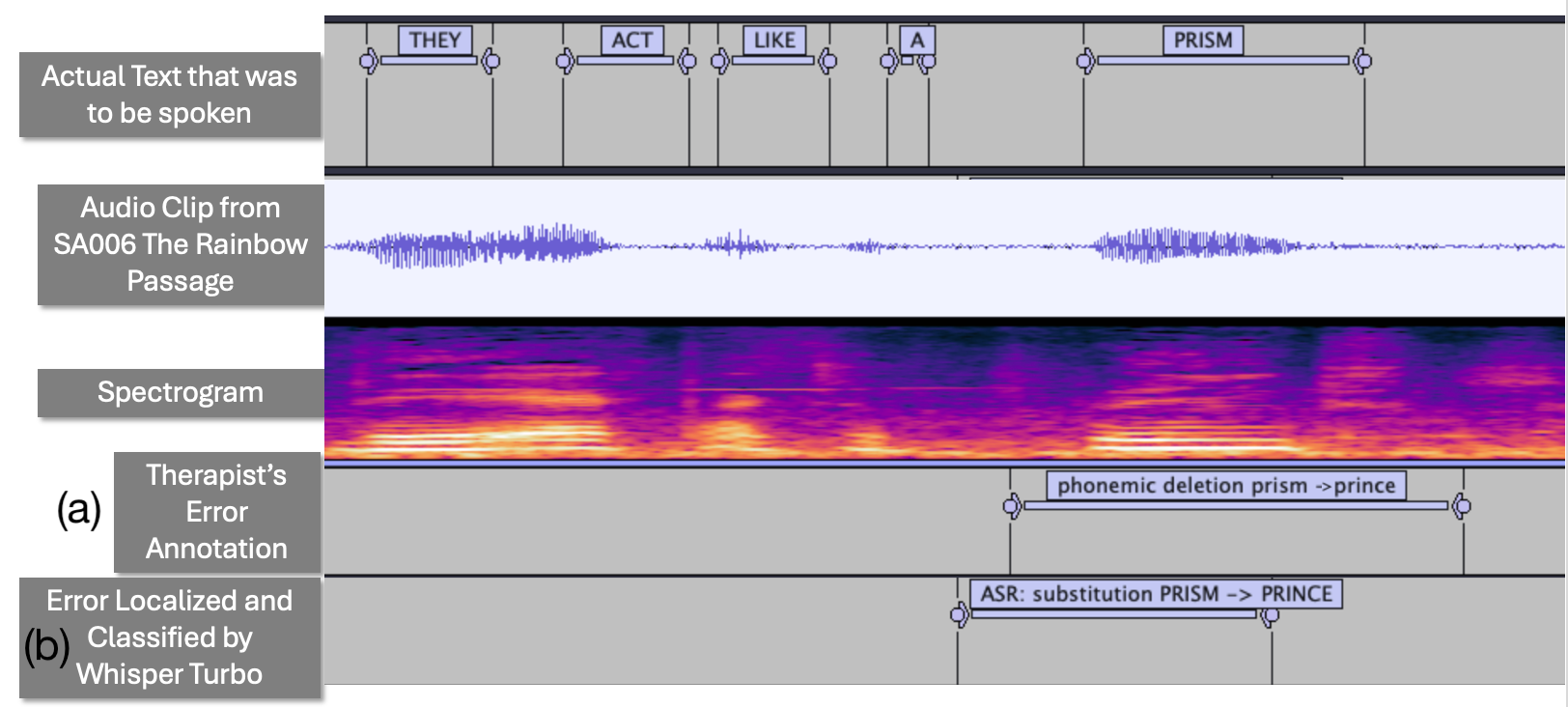}
    \vspace{-0.3cm}
    \caption{An example of annotation and the error derived from an ASR's output (Audio samples on webpage$^4$).}
    \label{fig:annotations}
    \vspace{-0.2cm}
\end{figure}

All annotations were made using Audacity\footnote{\url{https://www.audacityteam.org/}}, with temporally marked speech errors. Intelligibility was calculated as the \% of correctly produced words relative to the total words in each passage, representing the speech clarity perceived by the expert speech therapist, i.e., ~clarity score. Table \ref{tab:speakers} provides details of the speakers, including severity levels based on clarity scores.
\begin{table}[h!]
\centering
\vspace{-0.2cm}
\caption{Speaker Details. Total number of words in the two passages is 230. ``Speech Duration'' is the total recorded speech duration of each speaker. ``\#words in error'' are the total number of words that were marked as mispronounced region by the therapist. Clarity Score \% = 100*(1-\#words in error/230). Severity level based on clarity score: mild (80\% and above), moderate (50-80\%), and severe dysarthria (below 50\%).}
\label{tab:speakers}
\vspace{-0.2cm}
\renewcommand{\arraystretch}{1}
\setlength{\tabcolsep}{3pt} 
\resizebox{0.98\linewidth}{!}{\begin{tabular}{p{1cm}p{1cm}p{1cm}p{2cm}p{1.5cm}p{1.2cm}p{1.5cm}}
\hline
\textbf{ID} & \textbf{Gender}& \textbf{Age}&\textbf{Speech Duration (min)}& \textbf{\# Words in Error} & \textbf{Clarity Score \%} & \textbf{Severity Level}\\ \hline
SA001&M&71&2.08&63&72.6&Moderate\\\hline
SA002&M&60&2.21&52&77.4&Moderate\\\hline
SA003&M&47&2.88&6&97.3&Mild\\\hline
SA004&M&41&1.87&59&74.3&Moderate\\\hline
SA005&M&71&1.64&23&90.0&Mild\\\hline
SA006&F&69&2.05&45&80.4&Mild\\\hline
\end{tabular}}
\vspace{-0.2cm}
\end{table}
\\
\textbf{Mispronunciation Annotations: }
Each mispronounced region was annotated with a description label by the speech therapist, totaling 196 labels across the dataset. Each label consisted of the $\langle error~type\rangle$ followed by the $\langle exact~error \rangle$, consistent with standard practices in speech therapy. For example, the error description label ``\textit{word replacement find $\rightarrow$ found}'' consists of the $\langle error~type\rangle$ ``\textit{word replacement}'' and the $\langle exact~error \rangle$ ``\textit{find $\rightarrow$ found}'' meaning the word \textit{find} was replaced by the word \textit{found} (E.g.~Figure \ref{fig:annotations} (a)). 
We categorized these labels into five broad mispronunciation classes, which align with the characteristic patterns of dysarthric speech identified by the therapist: \\
\textbf{\textit{Substitution Errors}}:
    Replacing a phoneme or word with another, causing semantic or phonetic distortions. Eg: ``look'' pronounced as ``took'' or ``quivers'' as ``beer''. These errors reflect challenges in articulatory precision and phonological planning.\\
    \textbf{\textit{Deletion Errors}:} Omitting a phoneme or word, which disrupts word structure and meaning. E.g.: ``raindrops'' pronounced as ``raindops'' or ``likes'' as ``like''. They highlight reduced articulatory control and phonological simplification.\\
    \textbf{\textit{Insertion Errors}:} Adding an extra phoneme or word, i.e~introducing unintended elements. E.g.: ``nearly'' pronounced as ``nearlys'' or additional words like ``the'' or ``as''. These indicate difficulties in maintaining speech rhythm and lexical accuracy.\\
    \textbf{\textit{Repetition Errors}:} Repeating a phoneme or word, which disrupts fluency. E.g.: ``his'' pronounced as ``his his''. These reflect attempts at self-correction or motor planning challenges.\\
    \textbf{\textit{Prosodic Errors}:} Irregular or prolonged pauses, and strained voice, 
    which indicate impaired breath control and timing.

The dataset included 70 substitution errors, 39 deletion errors, 54 insertion errors, 13 repetition errors, and 20 prosodic errors. We used GPT-4-turbo with custom fine-tuned prompts\footnote{Webpage: \url{https://apps.ahlab.org/interspeech25_speechtherapy/}} to automatically map the descriptive labels into these five error categories. 
The mappings were manually verified and confirmed by two authors of this paper.

\vspace{-0.2cm}
\section{Speech Clarity Assessment Framework}
\begin{figure}
    \centering
    \vspace{-0.2cm}\includegraphics[width=0.8\linewidth]{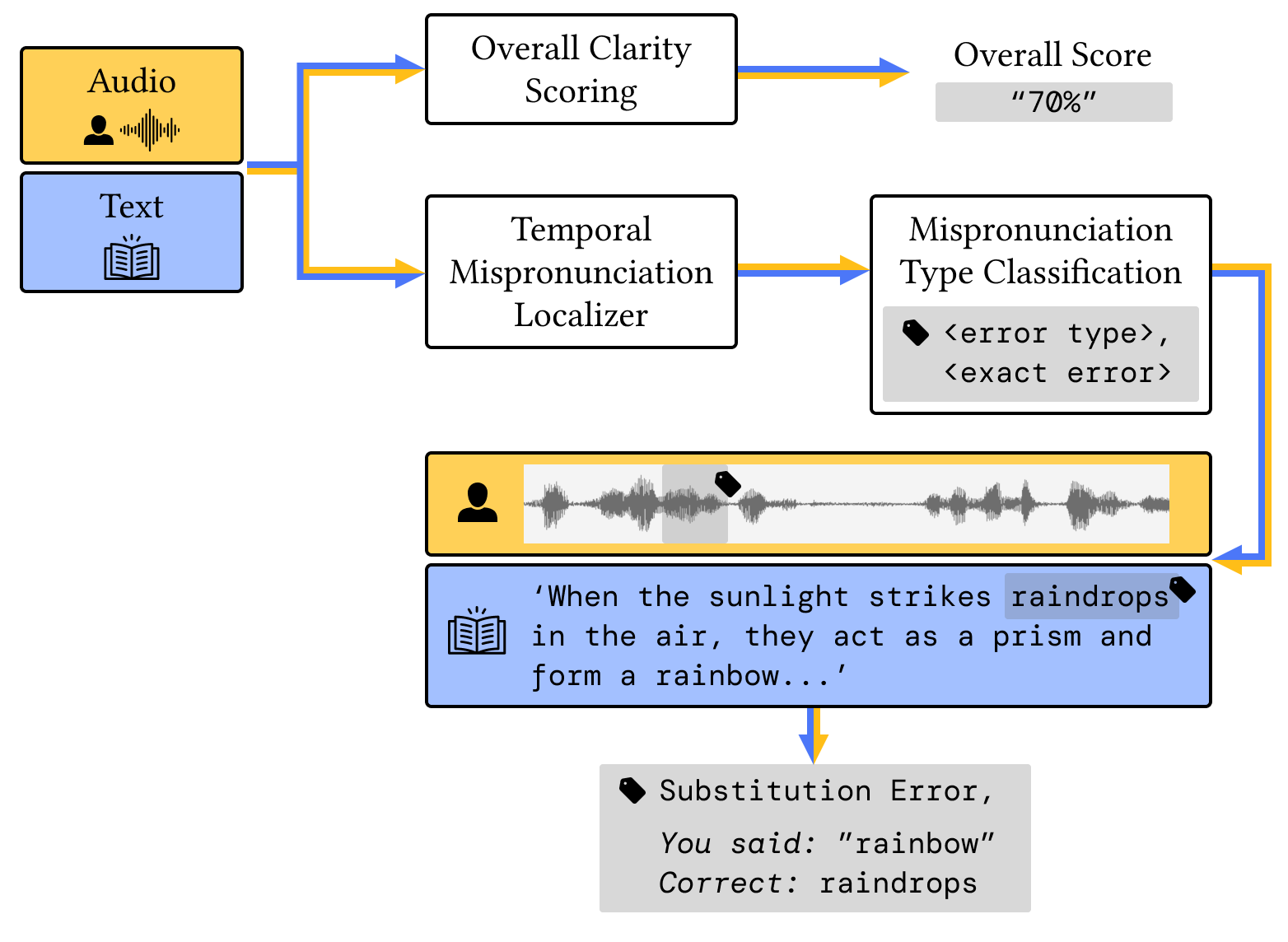}
    \vspace{-0.3cm}
    \caption{Framework for Speech Clarity Assessment}
    \label{fig:framework}
    \vspace{-0.7cm}
\end{figure}

We introduce a three-stage framework (Figure \ref{fig:framework}) for automated speech clarity assessment, designed for scenarios where a patient practices speaking a given text paragraph. In this setup, both the speech recording and the corresponding target text are available, enabling precise evaluation. Our framework is applied to our dysarthric speech dataset and analyzed using pre-trained ASR models to assess intelligibility, localize errors temporally, and classify mispronunciation types. Through this framework and dataset, we analyze the performance of existing ASR models in providing detailed feedback on dysarthric speech and derive insights to guide future work in this field. 
The stages of the framework are as follows:
\\
\textbf{Overall Clarity Scoring:} The first stage of our framework evaluates the overall clarity of each passage or utterance. This is done by comparing the transcript generated by state-of-the-art ASR models with the ground-truth text of the passages
resulting in a high-level indication of the speaker's overall speech clarity.
\\
\textbf{Temporal Mispronunciation Localizer:} In the second stage, we detect intelligibility-reduced regions by aligning ASR transcriptions with the text passage and extracting timestamps of the transcription regions in error. Using therapist annotations as ground truth, we assess ASR accuracy in localizing these errors and analyze which mispronunciation classes are better detected.
\\
\textbf{Mispronunciation Type Classification:} 
In the final stage, ASR output and reference text were converted into phoneme lists, with errors assessed using the Levenshtein distance. 
Word- and phoneme-level errors were distinguished using absolute ($\leq3$) and relative ($\leq0.6$) edit distance thresholds, selected empirically. 
The absolute distance ensures that phoneme-level classification is applied when the number of phoneme changes is three or fewer. The relative distance accounts for cases where the number of changes is proportional to the word's length. Examples are provided on our webpage.
Errors were classified into six types: phoneme- and word-level deletions, insertions, and substitutions. 
We evaluated these predicted error classes against the dataset’s mispronunciation labels, comparing both the $\langle error~type\rangle$ (mapped to eight classes, splitting word- and phone-level sub, del, ins) and $\langle exact~error\rangle$ against those from ASR. This stage provides a detailed characterization of error patterns, providing actionable insights for speech therapy.


\vspace{-0.1cm}
\section{Experiments and Results}
\subsection{Experimental Setup}
\vspace{-0.1cm}
\textbf{ASR Models:}
We employ widely used ASR models: wav2vec2-base and large, Whisper-tiny, base, small, medium, large, and turbo. Whisper models \cite{radford2022whisper} are transformer-based and optimized for robust, multilingual transcription, ranging from the lightweight tiny variant to the highly accurate large model, with turbo offering faster inference. The wav2vec2-based models \cite{baevski2020wav2vec} employ self-supervised learning on raw audio waveforms, with the large variant outperforming the base model. We use the whisper\_timestamped module \cite{lintoai2023whispertimestamped} to force align the ground-truth text with the audio to generate the start and end timestamps for each detected word. 
\\
\textbf{Baselines:}
We compare our results with previous work that evaluated off-the-shelf ASR models on dysarthric speech. Specifically, we consider \cite{leung2024training}, which tested Whisper models on the TORGO dataset without data augmentation. This enables a direct assessment of Whisper’s out-of-the-box performance on dysarthric speech in terms of overall speech clarity, using two datasets with comparable speaker diversity. 
For the second and third stages of our framework, we did not find comparable baselines in the literature for dysarthric speech.

\vspace{-0.2cm}
\subsection{Overall Clarity}
\begin{figure}
    \centering
    \includegraphics[width=0.95\linewidth]{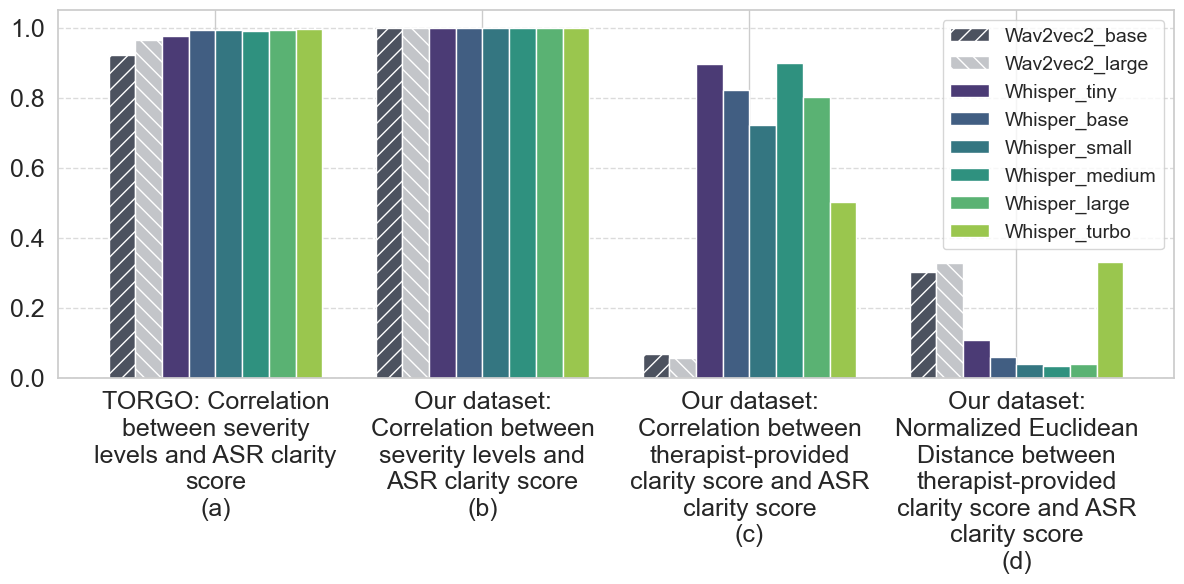}
    \vspace{-0.5cm}
    \caption{Comparison of the performance of the overall clarity scores generated different ASRs compared to the speaker severity levels and therapist-provided clarity scores.}\label{fig:overallclarity}
    \vspace{-0.7cm}
\end{figure}

\vspace{-0.2cm}
To evaluate the effectiveness of ASR models in assessing overall speech clarity, we analyze their ability to differentiate between speakers of different severity levels and compare their clarity scores against therapist-provided ratings. The overall clarity score derived from the ASR output  $=1-WER$, where WER (Word Error Rate) is the ratio of the sum of errors (substitutions, deletions, and insertions) from the ASR output to the total number of words in the ground-truth passage. 
To assess the trends, we compute the Pearson correlation between the ASR-derived clarity scores and the severity levels of the speakers in the datasets. Additionally, we measure the Pearson correlation and normalized Euclidean distance between the ASR-derived clarity scores and those from the therapists. \\
\textbf{Correlation with Dysarthria Severity Levels: }
We first examine how well ASR-predicted clarity scores align with the severity levels of speakers in both the TORGO dataset (four severity levels: 1, 2, 3, 4) and our dataset (two severity levels: 1, 2). The Pearson correlation between the average ASR clarity scores and their severity levels is high across both datasets (Figure \ref{fig:overallclarity} (a),(b)), indicating that ASR-based clarity scores reliably reflect dysarthria severity at a broad level. A similar trend can be observed in previous works \cite{troger2024automatic,leung2024training}, for instance, in \cite{leung2024training}, the correlation between the clarity scores from the Whisper medium model and the severity levels in the TORGO dataset is 0.98.\\
\textbf{Agreement with Therapist-Provided Clarity Scores: }
Unlike previous datasets, our dataset includes granular speech clarity ratings provided by therapists, allowing us to validate ASR-based clarity scores against expert assessments at a per-speaker level. We observe that medium and large Whisper models achieve the highest correlation with therapist-provided scores (Figure \ref{fig:overallclarity} (c)) while also demonstrating the lowest normalized Euclidean distance (Figure \ref{fig:overallclarity} (d)). This indicates that these models not only capture severity-level trends effectively but also generate clarity scores that closely align with expert judgments. However, in this case, the wav2vec2 models perform poorly, likely due to their reliance on frame-level acoustic features and monolingual training, which limits their ability to capture broader context and speaker and accent variability. In contrast, Whisper’s transformer-based architecture and multilingual training enhance context modeling and generalization across speech styles. Given that our dataset consists of long passages of English language speech with a local accent, Whisper’s ability to model context more effectively likely contributed to its better performance. Thus, we focus on Whisper models for subsequent analysis in this paper. 
\vspace{-0.3cm}
\subsection{Temporal Mispronunciation Localizer}
\begin{figure}
    \centering
    \includegraphics[width=0.9\linewidth]{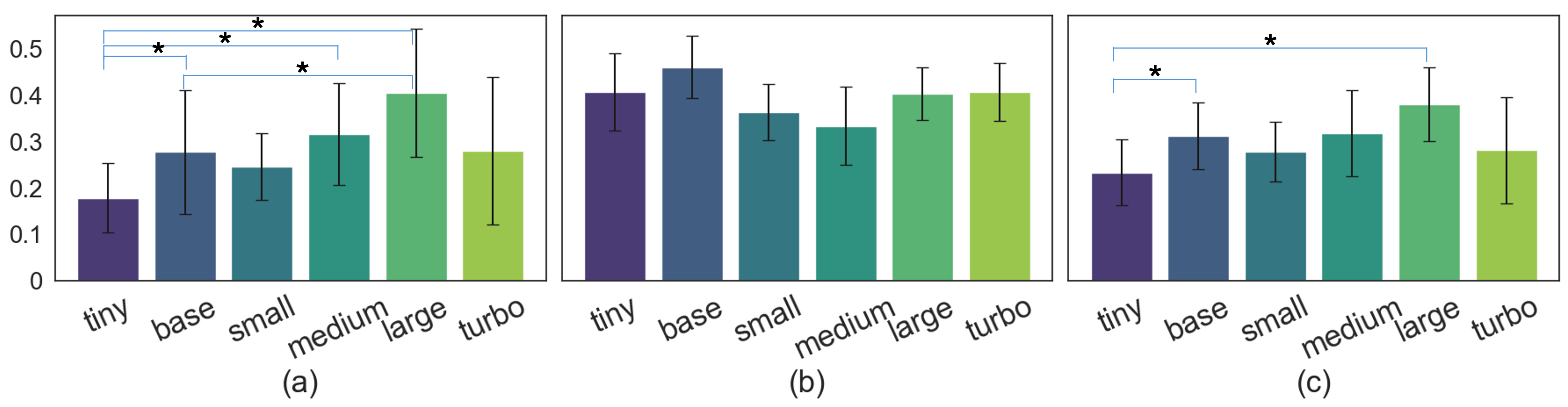}
    \vspace{-0.4cm}
    \caption{Comparison of the performance ($Mean\pm CI$) of the whisper models for temporal localization of mispronounced regions, in terms of (a) Precision, (b) Recall, and (c) F-score. Pairs showing statistically significant differences (Wilcoxon signed rank test $p<0.05$) are marked with *.}
    \label{fig:temporaloverview}
    \vspace{-0.4cm}
\end{figure}

\vspace{-0.2cm}
We use Precision, Recall, and F-score to measure the temporal accuracy of mispronunciation localization, such that: A \textit{true positive (TP)} is an ASR-detected error with a time overlap with any therapist-labeled error segment. A \textit{false positive (FP)} is an ASR-detected error that does not overlap with any therapist-labeled error. A \textit{false negative (FN)} is a therapist-labeled error that the ASR system failed to detect. Precision is $TP/(TP+FP)$, Recall is $TP/(TP+FN)$, and F-score is the harmonic mean of precision and recall. Figure \ref{fig:temporaloverview} presents the performance of different ASR models in localizing temporal mispronunciations in terms of precision, recall, and F-score corresponding to time overlaps between the ASR detected error and ground-truth from therapists. 

For precision, Shapiro-Wilk test indicated non-normal data distribution. Thus, we used the non-parametric Friedman test that showed a statistically significant difference across models ($\chi²(5) = 14.68, p = 0.012$). Post-hoc Wilcoxon tests identified significant differences ($p<0.05$) between Whisper-tiny and larger models (base, medium, large), and between Whisper-base and Whisper-large. Thus, Whisper-large notably outperformed smaller models in localizing mispronunciations with fewer false positives. For recall, there were no significant differences among the models, indicating similar detection capabilities.  For F-score, the data was not normally distributed, so we applied the Friedman test, that showed significant differences across models ($\chi²(5) = 12.41, p = 0.029$). Post-hoc Wilcoxon signed-rank tests identified significant differences ($p<0.05$) between Whisper-tiny and Whisper-base/large models. Overall, while all models perform similarly in recall, larger models achieve higher precision and F-scores, with Whisper-large being better at reducing false positives.

\begin{figure}
    \centering
    \includegraphics[width=0.9\linewidth]{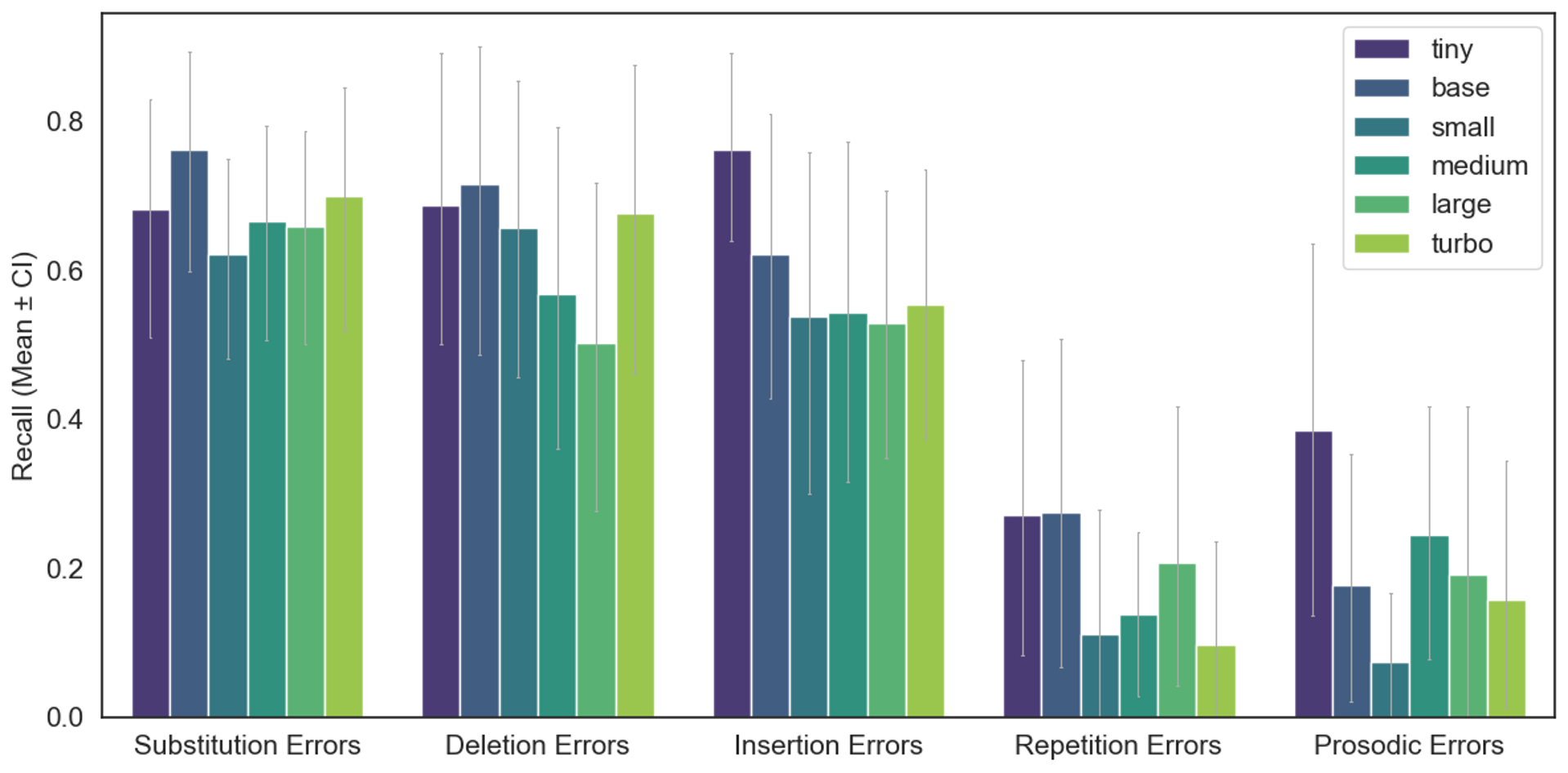}
    \vspace{-0.3cm}
    \caption{Accuracy of temporal localization of the ASR-detected errors (Recall) for  each mispronunciation class.}
    \label{fig:temporalgrouprecall}
    \vspace{-0.7cm}
\end{figure}

Recall performance across mispronunciation categories (Figure \ref{fig:temporalgrouprecall}) shows that substitution, deletion, and insertion errors are detected significantly more accurately ($0.64\pm0.3$) than repetition and prosodic errors ($0.19\pm0.3$) across all ASR models ($p<0.01$), 
indicating ASR models' difficulty in detecting prosodic and repetition errors, likely due to their subtle acoustic variations. There are no significant differences between ASRs within each error class.


\vspace{-0.3cm}
\subsection{Mispronunciation Classification}
\begin{figure}
    \centering
    \includegraphics[width=0.8\linewidth]{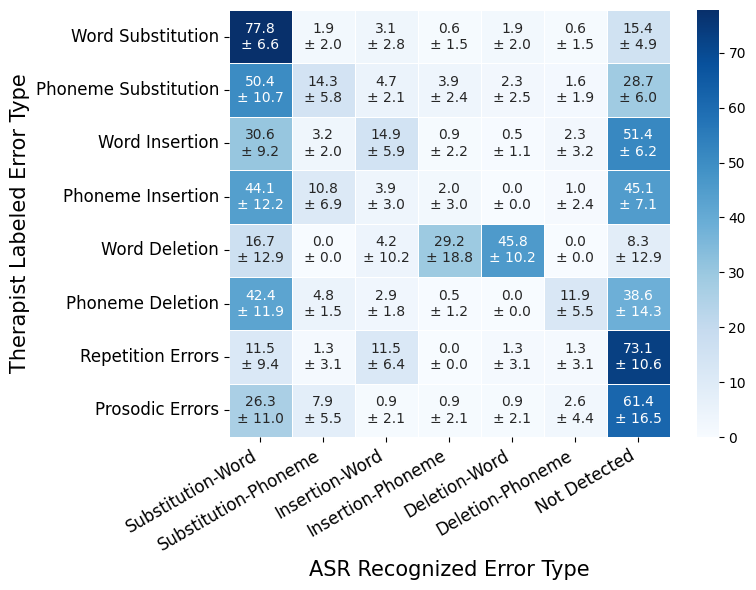}
    \vspace{-0.3cm}
    \caption{Mean\% ($\pm$Std) error-classification confusion matrix across ASRs (\% of total therapist's error count per class).}\label{fig:classification}
    \vspace{-0.6cm}
\end{figure}

\vspace{-0.2cm}
To evaluate the accuracy of mispronunciation classification, we compare word- and phone-level substitution, deletion, and insertion errors identified by the ASR models with the eight mispronunciation classes derived from therapists' annotations. The confusion matrix in Figure \ref{fig:classification} illustrates the \% of therapist-labeled errors classified into each ASR-recognized category, averaged across the six models. Substitution errors are consistently classified correctly across all models. Word deletion also has a high accuracy. 
However, phoneme-level errors are often misclassified as substitutions due to the post-recognition phoneme splitting process. For example, while a therapist may label 
\textit{swiftly$\rightarrow$swiftty} as phoneme deletion, the ASR recognizes the mispronounced \textit{swiftty} as \textit{fifty}, mapping it to the closest valid word. As a result, even after phoneme splitting, the error is classified as a word substitution rather than a phoneme deletion. 
Thus, direct phoneme recognition could improve classification accuracy. Additionally, repetition and prosodic errors remain largely undetected, consistent with temporal analysis findings.

Next, we analyze the true positives identified by the temporal mispronunciation localizer and manually verify whether the $\langle exact~error\rangle$ indicated by the speech therapist matches those derived from the ASR transcriptions. Often, therapists annotate mispronounced words as non-word utterances to accurately capture instances where patients with dysarthria produce non-words, e.g., \textit{prism $\rightarrow$ pris}, and \textit{swiftly $\rightarrow$ switty}. However, ASR models often interpret these mispronunciations as the closest-sounding real words, such as \textit{prism $\rightarrow$ prince}, and \textit{swiftly $\rightarrow$ fifty}. 
We considered such rhyming or phonetically similar words as correct matches with the therapist's exact error labels. This approach ensures a fair assessment of the ASR's ability to capture mispronunciations, even when the output diverges from the non-word annotations provided by therapists.

Across the ASR models, the average percentage of exact error matches with therapist annotations for temporally localized errors was 70.1\%$\pm$ 3.6\%. Substitution errors showed the highest match rates, with larger models achieving $\sim$80\% agreement. Deletion errors followed closely, with match rates improving as model size increased, while insertion errors had moderate accuracy. Repetition and prosodic errors showed high variability, but lower overall agreement compared to other error types. The high accuracy in exact error matching is promising, indicating potential for precise sub-word phonemic error identification. An example of localization and error assessment from Whisper turbo is shown in Figure \ref{fig:annotations}(b).

\vspace{-0.2cm}
\section{Conclusions and Future Work}
We presented a three-stage framework for automated mispronunciation evaluation in dysarthric speech, leveraging ASR models for (1) clarity scoring, (2) temporal mispronunciation localization, and (3) classification. Our expert-annotated dataset enabled systematic evaluation of ASR performance, with clarity scores showing strong alignment with dysarthria severity and therapist ratings. Larger Whisper models achieved higher precision and F-scores in temporal localization, though recall remained comparable across models. Classification was effective for substitution errors but less accurate for insertions, deletions, and prosodic errors due to ASR biases. Notably,  a 70.1\% exact match  between ASR-detected and therapist-labeled errors highlights the framework's potential for generating actionable feedback. These findings support the development of  scalable, clinically meaningful tools, that bridge ASR-based analysis and targeted therapeutic interventions. Future work will explore generalization by incorporating control speaker data and will focus on phoneme-level error detection, prosody-aware modeling, and dataset expansion to enhance robustness.

\bibliographystyle{IEEEtran}
\bibliography{mybib}

\end{document}